# Is Cloud Computing Steganography-proof?

Wojciech Mazurczyk, Krzysztof Szczypiorski
Institute of Telecommunications
Warsaw University of Technology
Warsaw, Poland
e-mail: {wmazurczyk, ksz}@tele.pw.edu.pl

*Abstract*— The paper focuses on characterisation of information hiding possibilities in Cloud Computing. After general introduction to cloud computing and its security we move to brief description of steganography. In particular we introduce classification of steganographic communication scenarios in cloud computing which is based on location of the steganograms receiver. These scenarios as well as the threats that steganographic methods can cause must be taken into account when designing secure cloud computing services.

## I. INTRODUCTION

Cloud computing is a very popular but still evolving paradigm. It allows customers of cloud providers to avoid start-up costs and/or reduce operating cost. It also increases their flexibility by immediately acquiring services and infrastructural resources when needed. NIST (National Institute of Standards and Technology) defines cloud computing as [1]: "a model for enabling ubiquitous, convenient, on-demand network access to a shared pool of configurable computing resources (e.g., networks, servers, storage, applications, and services) that can be rapidly provisioned and released with minimal management effort or service provider interaction". Key characteristics of cloud computing include: on-demand self-service, broad network access, resource pooling, rapid elasticity, and measured service. NIST defines also three service models in which cloud users control:

1. Only user-specific application configurations. It is called *SaaS* (Software as a Service),
2. Deployed applications and possibly application hosting environment configurations. It is called *PaaS* (Platform as a Service),
3. Everything (e.g. operating systems, storage, deployed applications) except the underlying cloud infrastructure. It is called *IaaS* (Infrastructure as a Service).

## II. SECURITY OF CLOUD COMPUTING

However, cloud computing uniqueness when compared with previous computing approaches launched new security, privacy and trust challenges from which most important are [2]:
1. Ensuring that only authorized parties have access to user data even if it is a service provider,
2. Sharing responsibility between cloud provider and its customer for security and privacy,
3. Providing secure multi-tenant environment (secure and efficient partitioning of virtualized and shared infrastructure among different customers).

Cloud Security Alliance defined seven most important threats for cloud computing [3] (only those who required it were commented):
- Threat #1: Abuse and Nefarious Use of Cloud Computing (different "activities" e.g. spamming, malware code, passwords cracking, DDoS attacks, botnet C&C etc.),
- Threat #2: Malicious Insiders,
- Threat #3: Data Loss or Leakage,
- Threat #4: Account or Service Hijacking,
- Threat #5: Insecure Interfaces and APIs.
- Threat #6: Shared Technology Issues – many underlying components of cloud computing services were not designed to offer strong isolation for multi-tenancy. Attackers focus on the operations of other cloud customers, and how to gain unauthorized access to their data,
- Threat #7: Unknown Risk Profile - ambiguous information about with who you will be sharing your infrastructure, in addition to logging data e.g. network intrusion logs, redirection attempts and/or successes etc.

Abovementioned threats can be divided into two groups. First, well-known threats group (Threats #1-5) which can be amplified by utilizing cloud computing environment. For example, attacks that used to be specific to a particular operating system in cloud computing become cross-platform. Second, cloud-specific threats group that exploit cloud computing characteristic features (Threats #6-7).

## III. STEGANOGRAPHY IN CLOUD COMPUTING

Steganographic techniques can be used to provide a perfect tool for data exfiltration, to enable network attacks or hidden communication among secret parties. The aim of these techniques is to hide secret data (steganograms) in the innocent looking carrier e.g. in normal transmissions of users. In ideal situation hidden data exchange cannot be detected by third parties. The best carrier for steganograms must possess two features: it should be popular i.e. usage such carrier should not be considered as an anomaly itself and modification of the carrier related to inserting the steganogram should not be "visible" to third party not aware of the steganographic procedure.

And how to find a carrier that would fill abovementioned requirements? In the Internet today we witness expansion of different, advanced Internet services from which more and more are migrating to abovementioned cloud computing services. The major cloud service providers are significantly investing in their infrastructure and in acquiring customers, big players list include: Google (Gmail, GoogleDoc), Microsoft (Azure), Amazon (Amazon Web Services), Cisco (WebEx). And these services utilise sometimes complex protocols and infrastructures to achieve their goals. Thus, they are perfect candidates for secret data carriers.

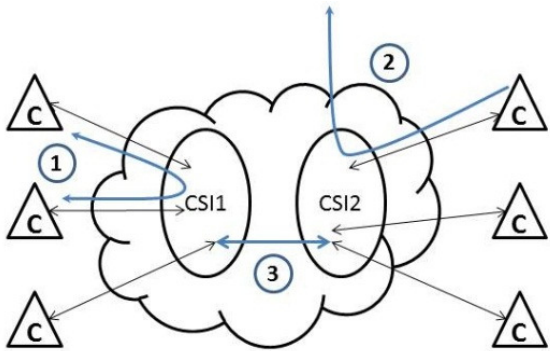

**Figure 1. Hidden communication scenarios for cloud computing environment**
**(CSI – Cloud Service Instance, C – Cloud Clients/Customers)**

We propose to consider three types of steganographic communication in cloud computing based on where the steganograms receiver (SR) is placed (Fig. 1). All these scenarios requires, of course, hiding intentions with which cloud services will be used:

(1) SR is situated in the same cloud instance as steganogram sender (SS). This case considers the scenario in which two cloud service users exchange confidential data for which at least one of them is unauthorised. Such communication includes utilisation of the so called *network steganography* – methods that as a carrier of secret data use network protocols (data units or the way they are exchanged), or relations between two or more protocols to enable hidden communication. This can lead to the realisation of the Threats #1-3.

(2) SR is situated outside the cloud service environment and cloud service capabilities are used to perform network attacks. In this case steganographic communication can be used e.g. to coordinate cloud botnet or malwares [4]. This type of communication includes also utilisation of network steganography. This can lead to the realisation of the Threats #1-3.

(3) SR is situated in the different cloud instance than SS. Such communication includes utilisation of the covert channel by using shared resources. Such solution were proposed based on data cross deduplication mechanism [5] or caches shared by services instances on the same physical machines [6]. This can lead to the realisation of the Threats #1-3 and 7.

## IV. CONCLUSIONS

To conclude, cloud computing as a carrier for secret communication is not very different from any other popular steganographic carriers e.g. like IP telephony [7]. The main novelty in this area when compared with known hidden data exchange opportunities is possibility of enabling secret communication between two instances of cloud services. Steganography should be treated for cloud computing environment as a rising threat to the network security as it is seen for typical networking ones.

Presented in this paper hidden communication scenarios as well as the threats that steganographic methods can cause must be taken into account when designing secure cloud computing services.

In order to minimize the potential threat of malicious use of steganography to public security effective steganalysis (detection) methods are needed. This requires in-depth understanding of the functionality of particular cloud service and the ways it can be used to enable hidden communication. Considering however variety and complexity of the cloud computing services there is not much hope that a universal and effective steganalysis method can be developed.

ACKNOWLEDGMENT

This work was partially supported by the Polish Ministry of Science and Higher Education under Grant: N517 071637 and IP2010 025470.